\documentclass[aps,pra,reprint,twocolumn,superscriptaddress]{revtex4-1}

\usepackage{graphicx} 
\usepackage{dcolumn}   
\usepackage{bm}       
\usepackage{amssymb}  
\usepackage{epstopdf}
\usepackage{amsmath}
\usepackage{calligra}
\usepackage{calrsfs}
\usepackage{hyperref}
\usepackage{xspace}
\usepackage{verbatim}
\usepackage{color}
\newcommand{\bra}[1]{\left\langle #1 \right|}
\newcommand{\ket}[1]{\left| #1 \right\rangle}
\newcommand{\up}{|\uparrow\rangle}
\newcommand{\updag}{\langle\uparrow|}
\newcommand{\down}{|\downarrow\rangle}
\newcommand{\downdag}{\langle\downarrow|}

\hypersetup{colorlinks=true,linkcolor=blue,citecolor=blue, filecolor=blue,urlcolor=blue,breaklinks=true}

\DeclareMathAlphabet{\mathcalligra}{T1}{calligra}{m}{n}

\begin{document}

\title{Macroscopic non-classical state preparation via post-selection}

\author{V\'{i}ctor Montenegro}
\email{vmontenegro@fis.puc.cl}
\affiliation{Instituto de F\'{i}sica, Pontificia Universidad Cat\'{o}lica de Chile,
Casilla 306, Santiago, Chile}

\author{Ra\'{u}l Coto}
\affiliation{Instituto de F\'{i}sica, Pontificia Universidad Cat\'{o}lica de Chile,
Casilla 306, Santiago, Chile}
\affiliation{Universidad Mayor, Avda. Alonso de C\'ordova 5495, Las Condes, Santiago, Chile}

\author{Vitalie Eremeev}
\affiliation{Facultad de Ingenier\'{i}a, Universidad Diego Portales, Santiago, Chile}

\author{Miguel Orszag}
\affiliation{Instituto de F\'{i}sica, Pontificia Universidad Cat\'{o}lica de Chile,
Casilla 306, Santiago, Chile}
\affiliation{Universidad Mayor, Avda. Alonso de C\'ordova 5495, Las Condes, Santiago, Chile}

\date{\today}

\begin{abstract}
Macroscopic quantum superposition (MQS) states are fundamental to test the classical-quantum boundary and present suitable candidates for quantum technologies. Although the preparation of such states have already been realized, the existing setups commonly consider external driving and resonant interactions, predominantly by considering Jaynes-Cummings and beam-splitter like interactions, as well as the non-linear radiation pressure interaction in cavity optomechanics. In contrast to previous works on the matter, we propose a feasible probabilistic scheme to generate a macroscopic mechanical qubit, as well as phononic Schr\"{o}dinger's cat states with no need of any energy exchange with the macroscopic mechanical oscillator. Essentially, we investigate an open dispersive spin-mechanical system in absence of any external driving under non-ideal conditions, such as the detrimental effects due to the oscillator and spin energy losses in a thermal bath at non-zero temperature. In our work, we show that the procedure to generate the mechanical qubit state is solely based on spin post-selection in the weak/moderate coupling regime. Finally, we demonstrate that the mechanical superposition is related to the amplification of the mean values of the mechanical quadratures as they maximize the quantum coherence. To the best of our knowledge, this physical mechanism has remained unexplored so far.
\end{abstract}

\maketitle

\section{Introduction} 

In the late 1920s, non-relativistic Quantum Mechanics (QM) was ultimately formulated to encompass the understanding of the microscopic and macroscopic world \cite{Wineland2013-Nobel, Friedman-Book}. Thus, for instance, there would be no objection to extend the quantum superposition principle to everyday life scale ---a very well-known conundrum established by E. Schr\"{o}dinger \cite{Schroedinger1935-Cat}. To date, macroscopic quantum superposition (MQS) appears not only to grasp fundamental aspects of QM \cite{DeMartini2012-QM, Romero-Isart2011}, but also as an excellent candidate for quantum technologies \cite{Bollinger1996-Precision, Riedel2010, Gross2010}.

Although, quantum superpositions at micro-scale have been widely realized (e.g. Refs. \cite{Donati1973-Electron, Zeilinger1988-Neutrons}), MQS states are more challenging to be achieved experimentally. This is because the large number of interacting particles and their interaction with its surroundings prevent the quantum behavior at macro-scale to emerge \cite{Zurek2003-Decoherence}. Despite this, MQS states have been demonstrated experimentally for some systems such as Josephson junctions \cite{Wilhelm2001-MQSJJ, vanderWal2000-MQSJJ, Friedman2000-MQSJJ, Clarke1988-Macro}, Cooper-pair boxes \cite{Nakamura1999-Cooper}, Bose-Einstein condensates (BECs) \cite{Ruostekoski1998-BECS, Huang2006-BECS}, Rydberg atoms \cite{Brune1996-Rydberg}, trapped ions \cite{Monroe1996-Ions}. On the other hand, quantum mechanical oscillators have attained increasing attention for MQS preparation due to notable experimental progress in the micro-fabrication of high-Q mechanical oscillators \cite{Schwab2005-Limitations, LaHaye2004-Limitations} in the quantum regime \cite{Rao2016-Cooling, OConnell2010-Ground, Teufel2011-GS, Peterson2016-GS, Yuan2015-GS}. Additionally, they can easily interact with an extensive range of physical systems, such as ultracold atomic BECs \cite{Treutlein2007-Condesate, Hunger2010-Condensate}, Cooper-pair boxes \cite{Steele2009-Dot, Armour2002-Cooper, Bose2006-Cooper}, opto-mechanical systems \cite{Scala2013-Levitated, Yin2013-Levitated}, etc. In particular, a superposition between two isolated states rises the possibility of having a long-lived mechanical qubit, which opens a window for quantum information technologies \cite{Leibfried2005-QIP, Gottesman2001-Encoding, Lee2011-Entangling}, quantum sensing \cite{Knobel2003-Sensing, Bollinger1996-Precision, Riedel2010, Gross2010}, as well as in the quantum communication field \cite{Reed2017}, e.g. as transfer interface in superconducting quantum circuits \cite{Palomaki2013-Transfer, Stannigel2010}, and they can serve as elements for modular quantum computation architectures \cite{Monroe2014, Narla2016}.

Stimulated by these, in the past few years, there have been remarkable advancements in the development of accurate quantum control and preparation of non-classical macroscopic mechanical states in different hybrid platforms as cavity/circuit QED \cite{Pirkkalainen2013-Cqed}, opto- and nano- mechanics \cite{Bose1999-Probe, Marshall2003-Superposition, Vitali2007-Mirror, Pepper2012-Superposition, Galland2014-Preparation, Aspelmeyer2014-Optomechanics, Carmele2014-Nano, Bochmann2013-Nano, Arcizet2011-NV, Wenchao2015-Macroscopic, Wenchao2015-Twomirrors, Milburn2016-Preparation}, trapped ions \cite{Blatt2008-Ions}, etc. Yet such schemes are based on resonant interactions, where the exchange of excitations between systems takes place. Moreover, non-linearities such as the radiation pressure in cavity opto-mechanics, the usage of external driving and interactions typically operating in the strong regime are required.

In this work, we present a scheme to prepare non-classical states of a macroscopic mechanical object. The protocol comprises a probabilistic qubit (0 and 1 phononic states) superposition, and the generation of mechanical Schr\"{o}dinger's cat state.  To realize this, we have considered an open spin-mechanical quantum system via a conditional displacement Hamiltonian in the dispersive regime without any need for adjusting resonances. Therefore, in comparison with previous works on the matter \cite{Heeres2015-Nonlinear, Krastanov2015-Pumping, Khan2016-Optomechanics}, our proposal does not rely on any non-linearity, energy exchange nor external pumping ---which might be an advantage for scalability purposes. Moreover, in contrast to cavity photons, spin systems exhibit both long coherence as well as depolarization times at room temperature, and also they can be easily prepared and readout \cite{Oort1988-Readout, Robledo2011-Readout}. Our probabilistic preparation protocol is uniquely based on two steps. Firstly, we weakly evolve the pre-selected spin-mechanical system for a time $t$, allowing us to truncate the oscillator Hilbert space up to a single phonon excitation. Subsequently, we then proceed to post-select the spin system, this step aims to prepare (probabilistically) any mechanical qubit superposition. Our results can be understood within the clear connection between the quantum coherence \cite{Baumgratz2014-Coherence} of the mechanics and the amplification of the position and momentum quadratures on average. 

This article is structured as follows. In Sec. \ref{sec:spin-mechanical dynamics}, we present the spin-mechanical model and we derive the mathematical condition to generate any macroscopic mechanical qubit in absence of any source of decoherence. In Sec. \ref{sec:master}, we have divided the discussion into two subsections. Firstly, we consider the open quantum case in presence of mechanical damping in a reservoir at zero temperature. Secondly, we take a closer experimental scenario by considering a full master equation, i.e we also include the spin relaxation and the pure dephasing terms. In addition to this, we consider spin-postselection inaccuracies. In Sec. \ref{sec:schrodinger}, we discuss how to generate mechanical Schr\"{o}dinger's cat states. In Sec. \ref{sec:aav} we give a very brief discussion on the connection between weak-measurements (AAV theory) and our protocol, a discussion which is extended in the Appendix section. Finally, in Sec. \ref{sec:conclusion}, we present the concluding remarks of our work.
 
\section{Macroscopic mechanical qubit preparation}\label{sec:spin-mechanical dynamics}
Let us commence by considering a spin qubit coupled dispersively to a mechanical oscillator. This elementary system is described in the interaction picture by ($\hbar = 1$)
\begin{equation}
\hat{H}_{\mathrm{int}} = \hat{b}^\dag \hat{b} - \lambda \hat{\sigma}_z (\hat{b}^\dag  +  \hat{b}),
\end{equation}

where $\lambda = \lambda_0/\omega_m$ is the scaled coupling strength, $\lambda_0$ the direct spin-mechanical coupling interaction and $\omega_m$ the oscillator frequency; $\hat{b}$ stands for the annihilation bosonic operator. To investigate the dynamics, we pre-select the spin as $\ket{\psi}_q(0) = 1/\sqrt{2}(\up + \down)$, and we initialize the mechanics in its ground state $\ket{\psi}_m(0) = \ket{0}$ \cite{Rao2016-Cooling, OConnell2010-Ground, Teufel2011-GS, Peterson2016-GS, Yuan2015-GS, GS-Justification}. In the following, we will show how a mechanical qubit can be generated via conditioned spin post-selection in the weak/moderate-coupling regime. To assess this, let us recast the spin-mechanical wave-function (in absence of any source of decoherence) as previously reported in \cite{Montenegro2014-NLO}; with $\eta = (1 - e^{-it})$
\begin{equation}
\ket{\psi(t)} = 1/\sqrt{2}(\up\ket{\lambda\eta}+\down\ket{-\lambda\eta}). 
\end{equation}

Firstly, for our procedure to succeed we require to have low mechanical quanta excitations, thus we proceed to truncate the mechanical coherent states $\ket{\pm\lambda\eta}$ up to their first phononic number state, i.e. $\ket{\pm\lambda\eta} \approx 1/\sqrt{1 + |\lambda\eta|^2}(\ket{0} \pm \lambda\eta\ket{1})$ ---an approximation valid when $|\lambda\eta| = \lambda\sqrt{2(1 - \cos t)} \ll 1$. This operational regime can be addressed, for instance, via magnetic coupling. There the interaction can be explicitly written as $\hbar \lambda_0 \approx \mu_B \partial B / \partial z \sqrt{\hbar / 2m\omega_m}$ \cite{Treutlein2014, Rabl2009}, where for a set of values of $\mu_B \sim 10^{-23}$ J/T (Bohr magneton), mass $m \sim 10^{-14}$ kg, mechanical frequency $\omega_m \sim 10^6$ Hz, and magnetic gradient between $10^4$ T/m $< \partial B / \partial z < 10^7$ T/m, the coupling can be reduced to $10^{-4} < \lambda < 10^{-1}$.

Subsequently, we post-select the spin with a general target state as $\ket{\psi_f} = \cos(\theta/2) \up + \sin(\theta/2)e^{i\phi} \down$. Thus the wave-function after the post-selection becomes
\begin{equation}
 \ket{\psi(t)}_m \approx \frac{1}{\mathcal{N}\sqrt{2(1+|\lambda\eta|^2)}}(\alpha_+\ket{0} + \lambda\eta\alpha_-\ket{1})\label{eq:mechanical_qubit}
\end{equation}

where, 
\begin{eqnarray}
\alpha_\pm &=& \cos(\theta/2) \pm e^{-i\phi}\sin(\theta/2),\\
\mathcal{N}^2&=&\frac{1+\sin\theta\cos\phi + |\lambda\eta|^2(1 - \sin\theta\cos\phi)}{2(1+|\lambda\eta|^2)}.
\end{eqnarray}

From (\ref{eq:mechanical_qubit}) one could easily notice that a combination of weak spin-mechanical coupling and spin post-selection can lead to a MQS. Particularly, for an equiprobable superposition, e.g., $|\bra{0}\psi(t)\rangle_m|^2 =|\bra{1}\psi(t)\rangle_m|^2 = 1/2$, we demand
\begin{equation}
 |\lambda\eta|^2(1-\sin\theta\cos\phi) = 1+\sin\theta\cos\phi. \label{eq:condition_super}
\end{equation}

The above equation stands as one of the main results of our work, as it relates the system dynamics ($\{\lambda,t\}$) with the required post-selection angles ($\{\theta_s, \phi_s\}$) to prepare the mechanics in a qubit state. It can be interpreted as following, if we let our system to evolve for a time $t$ (such as $\lambda\sqrt{2(1 - \cos t)} \ll 1$), then the mechanical qubit state will occur if and only if $\{\theta_s,\phi_s\}$ satisfies (\ref{eq:condition_super}), or vice-versa.

\section{A closer experimental realization of the mechanical qubit state}\label{sec:master}

The aim of the following Section is to study how our system would respond in presence of different sources of decoherence, and thus, investigate until which values we could accommodate our protocol before thermalization. To achive this, we have divided this Section into two, in (A) we study the system of interest uniquely considering oscillator energy losses in a reservoir at zero temperature. In subsection (B) we numerically solve a full master equation at non-zero temperature including also spin decoherences, as well as inaccuracies in the spin post-selection step.

\subsection{Open dynamics of the oscillator embedded in a reservoir at zero temperature}
The above derivation (\ref{eq:condition_super}) is restricted to a lossless evolution in a truncated Hilbert space. Nevertheless, it is of our best interest, to investigate the robustness of our scheme in a more realistic scenario. To model this we have solved the standard master equation for a reservoir at zero temperature. In this case, the master equation reads as
\begin{equation}
 \frac{d\hat{\rho}}{dt} = -i[\hat{H}_{\mathrm{int}},\hat{\rho}] + \frac{\gamma}{2}(2\hat{b}\hat{\rho}\hat{b}^{\dagger}-\hat{\rho}\hat{b}^{\dagger}\hat{b}-\hat{b}^{\dagger}\hat{b}\hat{\rho})
\end{equation}

where $\gamma$ (scaled by $\omega_m$) is the mechanical damping rate. Following the procedure described in \cite{Bose1997-Preparation}, one can analytically calculate the spin-mechanical density matrix as following $\hat{\rho} = \frac{1}{2}(\up\updag \otimes \hat{\Pi}_{++} + \up\downdag \otimes \hat{\Pi}_{+-} + \down\updag \otimes \hat{\Pi}_{-+} + \down\downdag \otimes \hat{\Pi}_{--})$, where we have defined
\begin{eqnarray}
 \hat{\Pi}_{\pm\mp} &=& e^{-\Gamma_{\pm\mp}^\gamma(t)} \ket{\beta_\pm(\gamma,t)}\bra{\beta_\mp(\gamma,t)},\\
 \hat{\Pi}_{\pm\pm} &=& \ket{\beta_\pm(\gamma,t)}\bra{\beta_\pm(\gamma,t)},\\
 \beta_\pm(\gamma,t) &=& \pm2 i \lambda \frac{(\gamma-2i)}{\gamma^2+4}\left(1-e^{-\frac{1}{2} (\gamma+2 i) t}\right),\\
 \Gamma_{\pm\mp}^{\gamma}(t) &=& -\frac{\gamma}{2} \int_0^t |\beta_\pm(\gamma,t') - \beta_\mp(\gamma,t')|^2 dt'.
\end{eqnarray}

Under mechanical damping, the normalized mechanical state after the spin post-selection becomes
\begin{equation} 
 \hat{\rho}_m = \frac{\cos^2\frac{\theta}{2}\hat{\Pi}_{++} + \sin^2\frac{\theta}{2}\hat{\Pi}_{--}+\frac{\sin\theta}{2}(e^{i\phi}\hat{\Pi}_{+-} + h.c)}{2\mathcal{P}(\theta,\phi)},\label{mechanicalstate1}
\end{equation}

with
\begin{eqnarray}
\nonumber \mathcal{P}(\theta,\phi) &=& \frac{1 + \sin\theta\mathrm{Re}[e^{-D^\gamma_{+-}(t)}e^{i\phi}\bra{\beta_-(\gamma,t)}\beta_+(\gamma,t)\rangle]}{2}\\
\end{eqnarray}

being the probability to post-select the spin.

To exhibit the mechanical superposition, we calculate the phonon probability distribution $(\mathrm{Pr}(n) = \bra{n}\hat{\rho}_m\ket{n})$ obtained from Eq. (\ref{mechanicalstate1})
\begin{equation}
 \mathrm{Pr(n)} = \frac{4^n e^{-4 c_1\lambda^2} (c_1\lambda^2)^n}{2\mathcal{P}(\theta,\phi)n!}[1+\sin \theta\cos\phi e^{c_2\lambda^2} (-1)^n],\label{prob}
\end{equation}

where (for $t=\pi$) we have defined
\begin{eqnarray}
c_1 &=& \frac{(e^{-\frac{\pi\gamma}{2}}+1)^2}{\gamma^2+4},\\
\nonumber c_2 &=& \frac{8 e^{-\pi\gamma}}{(\gamma^2 + 4)^2} \\
\nonumber &\times&(4 e^{\frac{\pi\gamma}{2}} \gamma^2+\gamma^2-e^{\pi\gamma}(-3\gamma^2+\pi(\gamma^2+4)\gamma+4)+4)\\
\end{eqnarray}

\begin{figure}
 \centering \includegraphics[scale = 0.55]{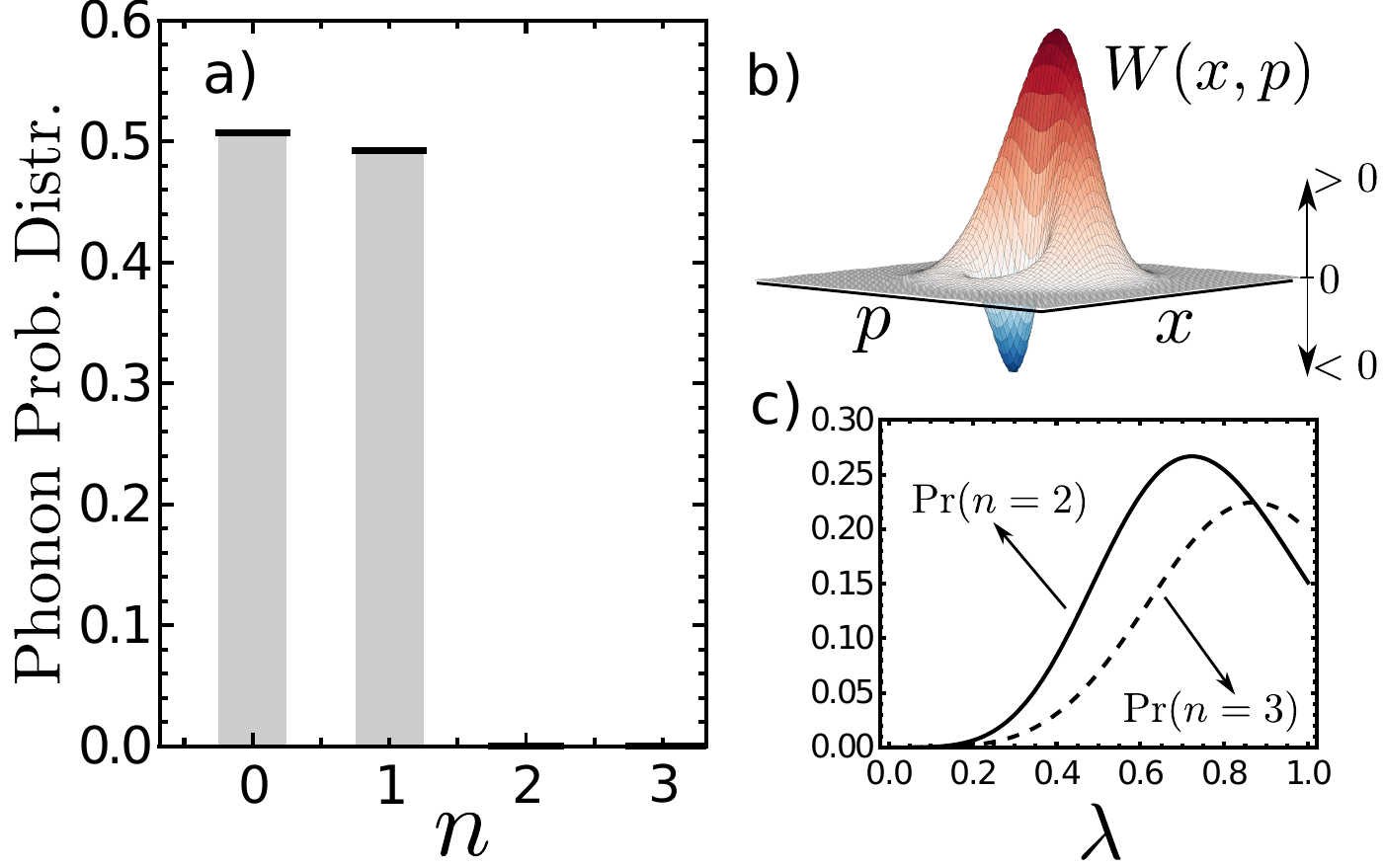}
 \caption{(Color online) Generation of a macroscopic mechanical qubit. (a) Phonon probability distribution as in (\ref{prob}). In (b) we illustrate the Wigner function for the case illustrated in (a). In (c) we calculate $\mathrm{Pr}(n)$ for $n=2$ and $n=3$ as in Eq. (\ref{prob}). As seen, for $\lambda \leq 0.25$ the mechanical qubit preparation remains still robust. Other parameters are $t=\pi, \gamma=10^{-2}, \lambda=0.1$, and $\{\theta_s,\phi_s\}$ are solution from (\ref{eq:condition_super}).}
 \label{qubit}
\end{figure}

A reasonable approximation satisfying $\mathrm{Pr}(n=0) = \mathrm{Pr}(n=1) \approx 1/2$ is to consider the lossless solution from Eq. (\ref{eq:condition_super}) (for $\gamma \ll 1$), and hence, to obtain the set of values $\{\theta_s,\phi_s\}$ from (\ref{eq:condition_super}). In Fig. \ref{qubit}-a we illustrate $\mathrm{Pr}(\mathrm{n})$ as in (\ref{prob}), giving us a very close equal probability distribution for $n=0$ and $n=1$.

As known, $\mathrm{Pr}(n=0) = \mathrm{Pr}(n=1) \approx 1/2$ are not sufficient conditions for quantum superposition by themselves, as fully mixed classical states can have same probabilities. Hence, to evidence the quantumness of our preparation, in Fig. \ref{qubit}-b, we have numerically computed the Wigner quasi-probability distribution defined as $W(x,p) = \frac{1}{\pi}\int_{-\infty}^\infty\left<x + x'\left| \hat{\rho}_{m}\right| x-x' \right>e^{-2ipx'}dx'$ \cite{Johansson2013-Qutip}, the true quantum nature arises as a consequence of the considerable negative part of $W(x,p)$.

Furthermore, one may wonder whether we can set an upper bound of the weak coupling condition $|\lambda\eta| \ll 1$. Certainly, the above ensures that the Hilbert space is properly truncated up to just one single phonon. However, we can fine-tune this assumption up to $\lambda \leq 0.25$, at the cost of having $\mathrm{Pr}(n=2) \approx$ 1.4\% (see Fig. \ref{qubit}-c). The benefits of having stronger $\lambda$ are to increase the post-selection outcome probability $\mathcal{P}(\theta,\phi)$, and also as $\lambda$ increases, then the qubit superposition becomes less susceptible to $\{\theta_s, \phi_s\}$ fluctuations (as we will study in the next Section). Suitable mechanical qubit candidates are found to be in the range of $0.05 < \lambda < 0.2$ with an average spin post-selection probability of $2\% < \mathcal{P}_{\mathrm{av}} <  24\%$.

The generated qubit in our work can be understood in the context of quantum coherence. As known, in quantum optics, it is well established that two or more quantum states of a single mode can interfere with themselves if they have non-zero coherence. Recently, in the domain of the quantum information science, this feature have also been demonstrated of being related with the amount of quantum entanglement \cite{Streltsov2015-Coherence}. For a given N$\times$N matrix $\hat{\rho} = \sum_{\{i,j\}=0}^N\rho_{i,j}\ket{i}\bra{j}$, the quantum coherence is defined as $\mathcal{C} = \sum_{\{i\neq j\}=0}^N|\rho_{i,j}|$ \cite{Baumgratz2014-Coherence}. In particular, for a 2$\times$2 matrix, $\mathcal{C}$ is reduced to (for $[\hat{x},\hat{p}]=\frac{i}{2}$)
\begin{equation}
 \mathcal{C} = |\bra{0} \hat{\rho}_m \ket{1}| + |\bra{1} \hat{\rho}_m \ket{0}|  = 2\sqrt{\langle \hat{x} \rangle^2 + \langle \hat{p} \rangle^2}. \label{eq:coherence}
\end{equation}

From the above expression, it is straightforward to obtain the maximum value $\mathcal{C}_{\mathrm{max}} = 1$ for the qubit case. Moreover, it relates the quantum coherence with the mechanical properties that we can, in principle, extract from an experiment. In addition to this, Eq. (\ref{eq:coherence}) shows the impossibility to generate a mechanical qubit superposition by its quantum evolution alone (i.e. by tracing out the spin state), as the expectation values are always zero in this case. Also, a post-selection on the $\hat{\sigma}_z$ eigenstates will not reach the required amount of coherence to generate a qubit superposition, as $\langle \pm2\lambda| \hat{x}|\pm2\lambda\rangle = \pm2\lambda, \langle \pm2\lambda|\hat{p}|\pm2\lambda\rangle=0$.

\begin{figure}
 \centering \includegraphics[scale = 0.6]{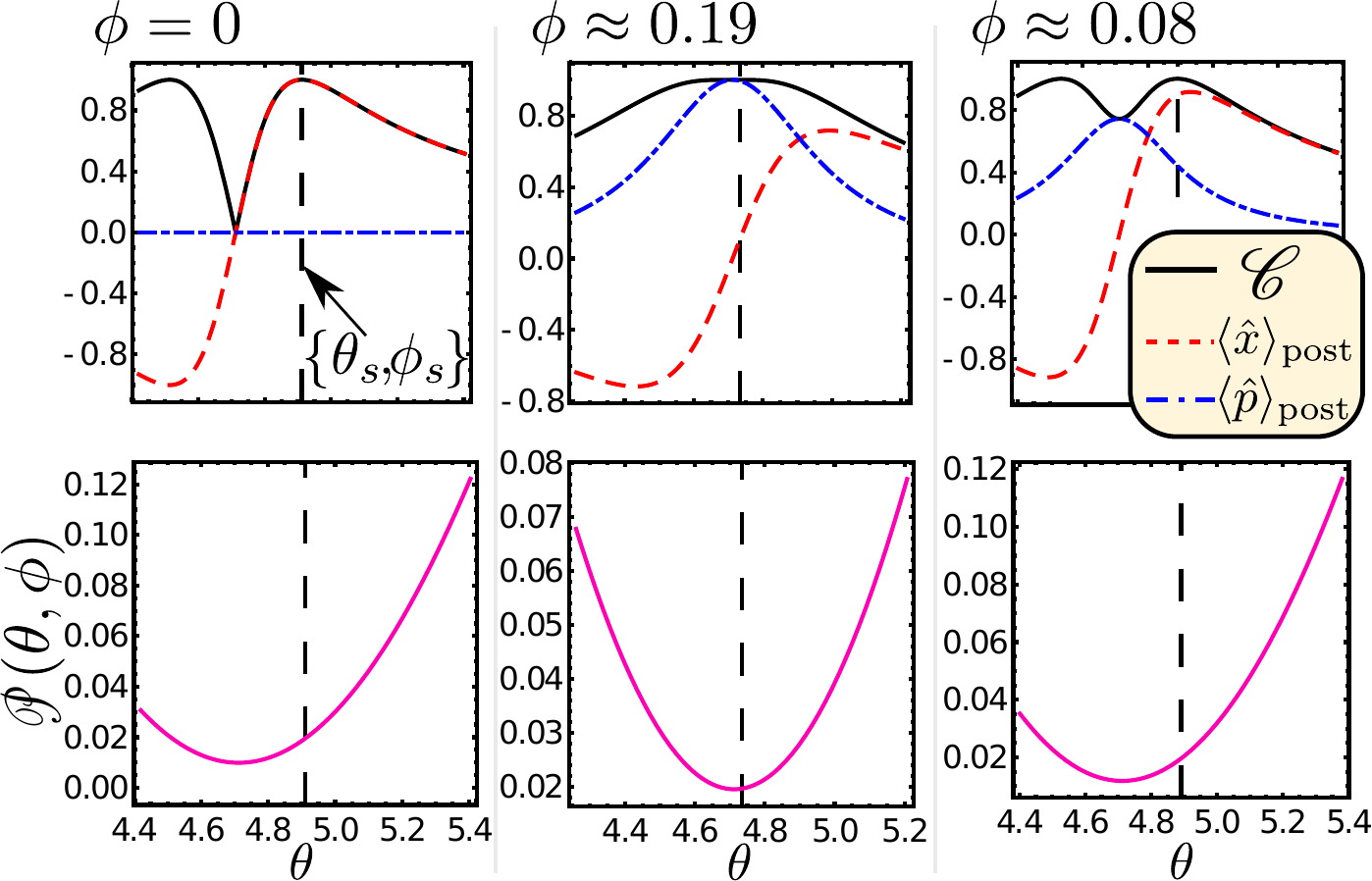}
 \caption{(Color online) (Top panel) Quadratures on average after post-selection and the mechanical coherence as in Eq. (\ref{eq:coherence}) for different $\phi_s$. The superposition is achieved when coherence is maximal for a set of post-selected angles $\{\theta_s,\phi_s\}$ satisfying (\ref{eq:condition_super}) (vertical dashed line). (Bottom panel) Outcome post-selected probability $\mathcal{P}$ for the same angle window. Other parameters are $t=\pi,\gamma=0.01,\lambda=0.05$.}\label{fig:coherence}
\end{figure}

In the top panel of Fig. \ref{fig:coherence}, we plot the mechanical coherence, the mean values of the position and the momentum quadratures. There, we show three cases corresponding to three different $\phi_s$ angles, where the coherence reaches approximately its maximum value ($t = \pi$ and $\gamma = 0.01$) for its corresponding $\theta_s$, and it coincides with i) the maximum of $\langle \hat{x} \rangle_{\mathrm{post}}$ ($\langle \hat{p} \rangle_{\mathrm{post}} = 0$), ii) the maximum of $\langle \hat{p} \rangle_{\mathrm{post}}$ ($\langle \hat{x} \rangle_{\mathrm{post}} = 0$), and iii) a combination of both, respectively. This is also confirmed from Eq. (\ref{eq:coherence}). In the bottom panel of Fig. \ref{fig:coherence} we show their corresponding probabilities, where for $\lambda = 0.05$ then $\mathcal{P}_{\mathrm{av}} \approx 2\%$.

\subsection{Open spin-mechanical dynamics in a non-zero temperature reservoir and spin post-selection inaccuracies}

As shown above, we solved the dynamics of the open quantum system uniquely involving the mechanical energy losses within a thermal reservoir at zero temperature. Although this might be considered as a first step approximation towards realistic experimental scenarios, it is further required to take into account unavoidable detrimental effects due to the spin decoherence and embedded in a thermal phonon reservoir at non-zero temperature. To model this case, we proceed to solve numerically a more general master equation given by:

\begin{eqnarray}
\nonumber \frac{d\hat{\rho}}{dt} &=& -i[\hat{H}_\mathrm{int},\hat{\rho}] + \gamma(1+\overline{n}_{m})\mathcal{D}[\hat{b}] + \gamma\overline{n}_{m}\mathcal{D}[\hat{b}^{\dagger}] \\
&+& \Gamma(1+\overline{n}_{q})\mathcal{D}[\hat{\sigma}^-] + \Gamma\overline{n}_{q}\mathcal{D}[\hat{\sigma}^+] + \frac{\gamma_\phi}{2}\mathcal{D}[\hat{\sigma}_z]\label{master2}
\end{eqnarray}

where $\hat{H}_\mathrm{int} = \hat{b}^\dagger\hat{b} - \lambda \hat{\sigma}_z (\hat{b}^\dagger  +  \hat{b})$ and
\begin{equation}
  \mathcal{D}\left[\hat{O}\right] = \frac{1}{2}\left(
  2\hat{O}\hat{\rho}\hat{O}^{\dagger}-
  \hat{\rho}\hat{O}^{\dagger}\hat{O}-\hat{O}^{\dagger}\hat{O}\hat{\rho}\right)
\end{equation}

corresponds to the Lindblad term. Also, in the equation above $\overline{n} = \left[ \mathrm{exp}(\hbar \omega / k_B T) - 1 \right]^{-1}$ is the Planck distribution, being $k_B$ the Boltzmann's constant and $T$ the corresponding temperature of the thermal phonon reservoir. Furthermore, the scaled (by the mechanical frequency $\omega_m$) quantities $\{\gamma, \Gamma, \gamma_\phi\}$ are the mechanical damping, spin relaxation, and the spin pure dephasing rates, respectively. For simplicity we have considered $\overline{n}_{m} = \overline{n}_{q} = 10$ throughout our numerics.

To quantify the robustness of our setup we make use of the fidelity, where for two quantum states $\{\varrho_1, \varrho_2\}$ is defined as $\mathrm{Tr}\left[\sqrt{ \sqrt{\varrho_1} \varrho_2 \sqrt{\varrho_1}} \right]$. In particular, as our target state is a pure state $\varrho_1 = \ket{\psi_m}\bra{\psi_m}$, where $\ket{\psi_m} = 1/\sqrt{2} (\ket{0} +  \ket{1})$ \cite{explanationplussign} is the  mechanical qubit in absence of any source of decoherence. The fidelity then reduces to:
\begin{equation}
 fidelity = \sqrt{ \bra{\psi_m} \varrho_2 \ket{\psi_m}}.
\end{equation}

\begin{figure}[t]
 \includegraphics[scale = 0.4]{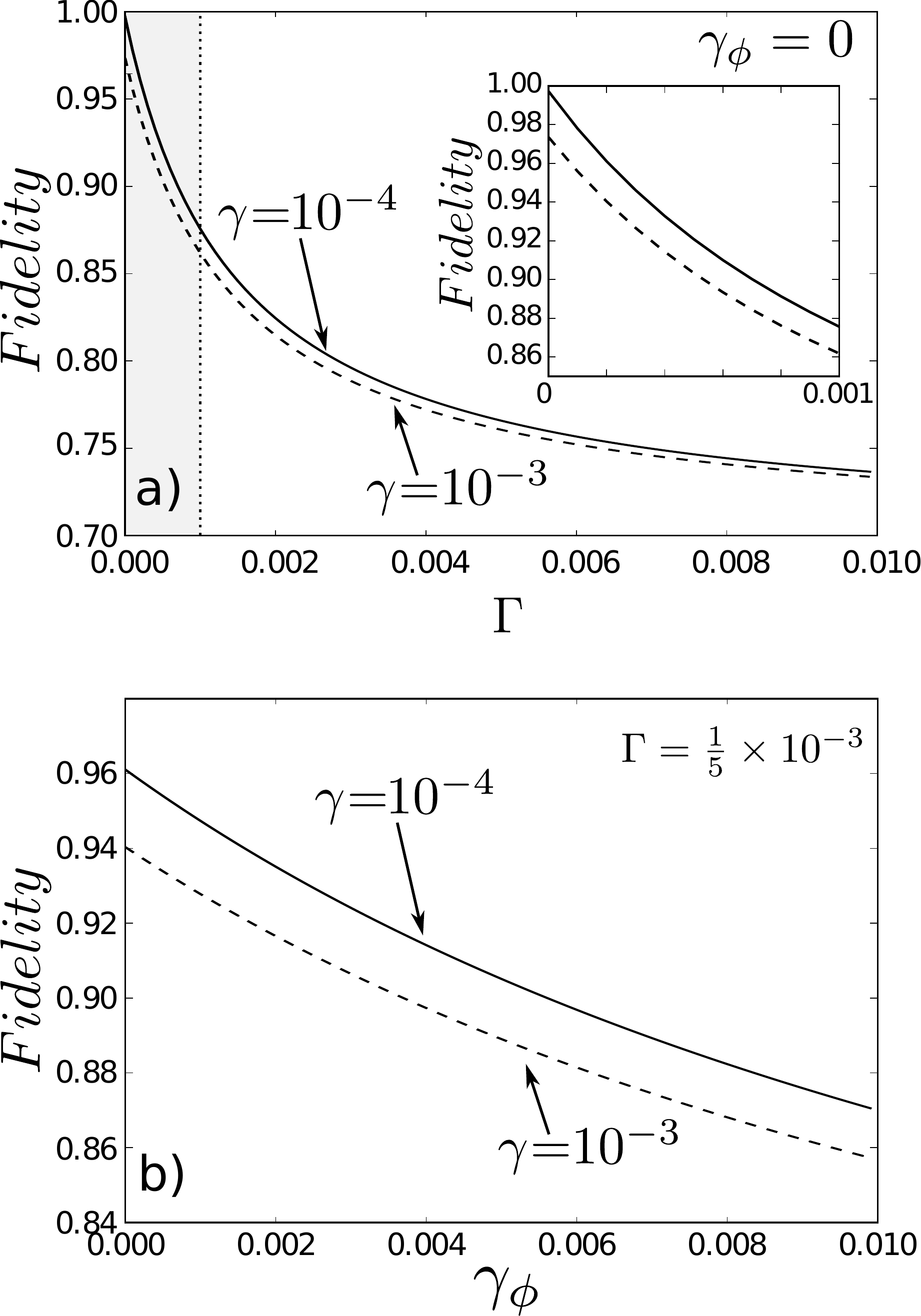}
 \caption{The figure illustrates the open quantum dynamics for two different values of the mechanical rate ($\gamma$). In (a) we solved Eq. (\ref{master2}) in absence of the spin pure dephasing rate ($\gamma_\phi = 0$) as a function of the spin relaxation rate $\Gamma$. The inset figure corresponds to the shaded region of the main plot. In panel (b) we show the evolution of the whole master equation shown in Eq. (\ref{master2}). Other values are ; $t =  \pi, \overline{n}_m = \overline{n}_q = 10, \lambda = 0.05$, and we have post-selected the spin as $\phi = 0$ and $\theta$ according to Eq. (\ref{eq:condition_super}).} \label{fig:master_equation}
\end{figure}

Now, we have all the ingredients to explore the limitations of our mechanical qubit state preparation under a closer experimental realization. Firstly, let us commence by studying our protocol in absence of spin dephasing, i.e., $\gamma_\phi = 0$ in (\ref{master2}). In Fig. \ref{fig:master_equation}-a we show the open evolution as a function of the spin relaxation rate $\Gamma$ for two values of the mechanical damping rate $\gamma$. As seen from the figure, in order to have a fidelity above 0.85, we require to have a value of $\Gamma < 10^{-3}$ (shaded region in Fig. \ref{fig:master_equation}-a). In Fig. \ref{fig:master_equation}-b we study the quantum open dynamics for $\gamma_\phi \neq 0$. To achieve this, we fixed a ``good enough'' spin relaxation value found in the previous (shaded region) figure, and we proceed to plot the fidelity as a function of the dephasing rate $\gamma_\phi$ for some values of $\{\gamma, \Gamma\}$. As shown in the bottom panel of Fig. \ref{fig:master_equation}, the fidelity can reach values up to above $0.86$ even under the full spin-mechanical open evolution.

Moreover, we have to recall that throughout our manuscript we have illustrated the operational coupling regime by considering frequencies of the mechanical oscillator in the range of $\mathrm{MHz}$, therefore with a phonon number occupancy of $\overline{n}_m \approx 10$ for a temperature on the order of milli-Kelvin. Nevertheless, it is important to notice that even though we considered $\overline{n}_m \approx 10$, the occupation number of the phonons can indeed be relaxed to higher values. Naturally, this statement will depend on the mechanical quality factor (Q=$\gamma^{-1}$) considered by us. As seen from Fig. \ref{fig:master_equation}, we considered values of Q=$10^{3}$ and Q=$10^{4}$ with $\overline{n}_m \approx 10$. Here, we notice that the upward $\gamma(1+\overline{n}_{m})$ and downward $\gamma\overline{n}_{m}$ energy transitions in the Lindbladian are of the order of $\gamma(1+\overline{n}_{m}) \approx \gamma\overline{n}_{m} \sim 10^{-2}$. Nowadays, mechanical oscillators can be prepared with Q=$10^5$ \cite{Jaehne2008}, impliying that our setup would still work considering a thermal bath with $\overline{n}_m \approx 10^2$. In fact, we have numerically verified this, i.e our non-classical mechanical preparation can still be generated for a set of values of $\{\overline{n}_m, \gamma, \Gamma, \gamma_\phi\} = \{10^2, 10^{-5}, 10^{-4}, 10^{-3}\}$, giving us $\mathrm{Pr}(n=0) = 0.518 \pm 0.024$ and $\mathrm{Pr}(n=1) = 0.447 \pm 0.022$.

\begin{figure}[t]
 \includegraphics[scale = 0.3]{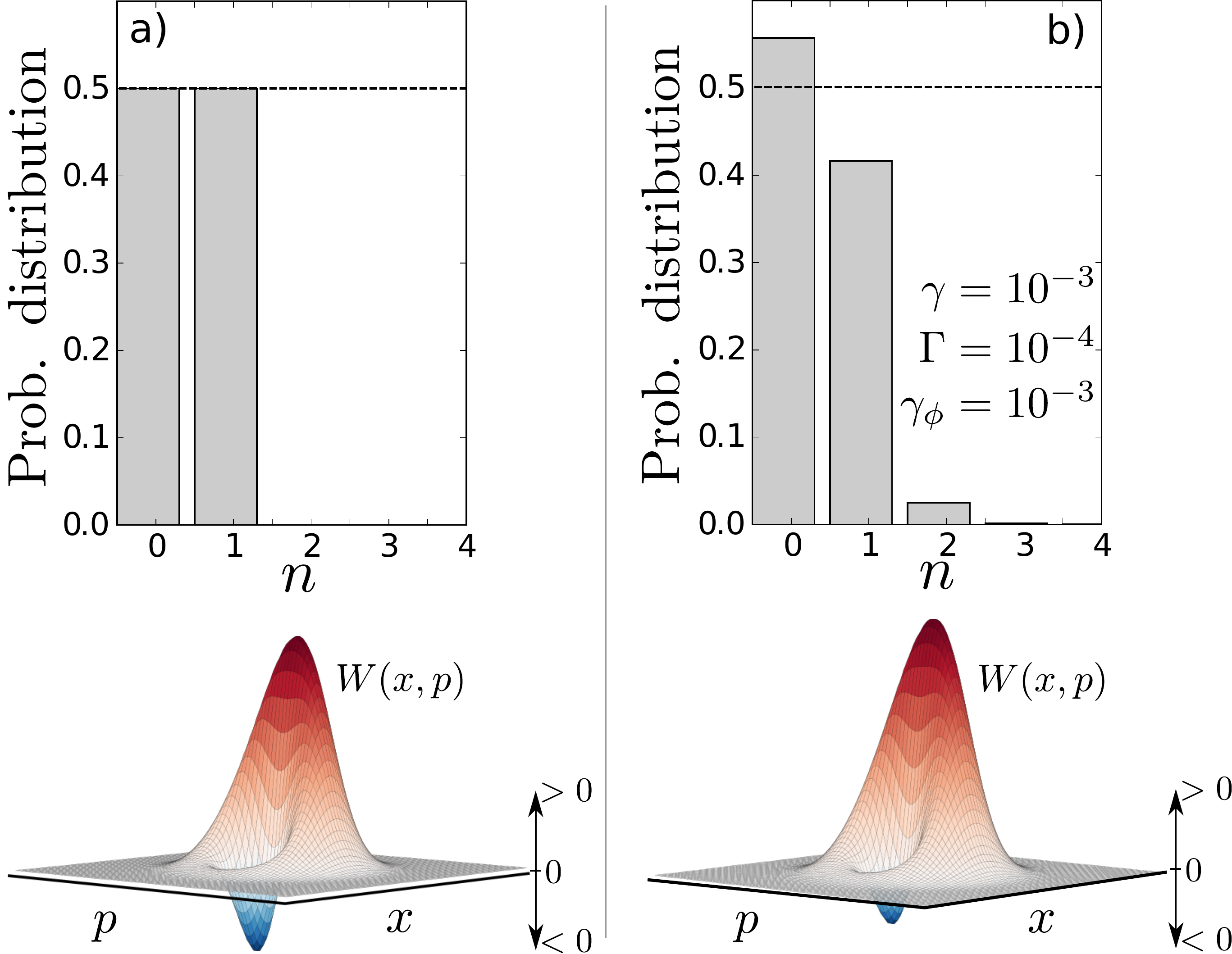}
 \caption{(Color online) Probability number and Wigner distributions of (a) the mechanical qubit preparation in absence of any source of decoherence, and (b) by solving the full master equation shown in Eq. (\ref{master2}). Our protocol can be accommodated for decoherence values of the order of $\{\gamma, \Gamma, \gamma_\phi\} = \{10^{-3}, 10^{-4}, 10^{-3}\}$. Other values were taken as in Fig. \ref{fig:master_equation}} \label{fig:prob}
\end{figure}

Following with a similar analysis discussed before, in order to exhibit the true quantumness of our mechanical state preparation, we proceed to compute the probability number as well as the Wigner quasi-probability distribution. In Fig. \ref{fig:prob}-a we show the state preparation under unitary (ideal) evolution, whereas in Fig. \ref{fig:prob}-b we consider the mechanical qubit under both spin and mechanical decoherence embedded in a non-zero thermal reservoir. There, the prominent negative area of the Wigner distribution exhibits the non-classical feature of our mechanical state preparation. We can conclude that, in principle, our protocol might be accommodated for a set of values of the order of $\{\gamma, \Gamma, \gamma_\phi\} \approx \{10^{-3}, 10^{-4}, 10^{-3}\}$ \cite{Aspelmeyer2014-Optomechanics, Bar-Gill2013}.

\begin{figure}[t]
 \includegraphics[scale = 0.28]{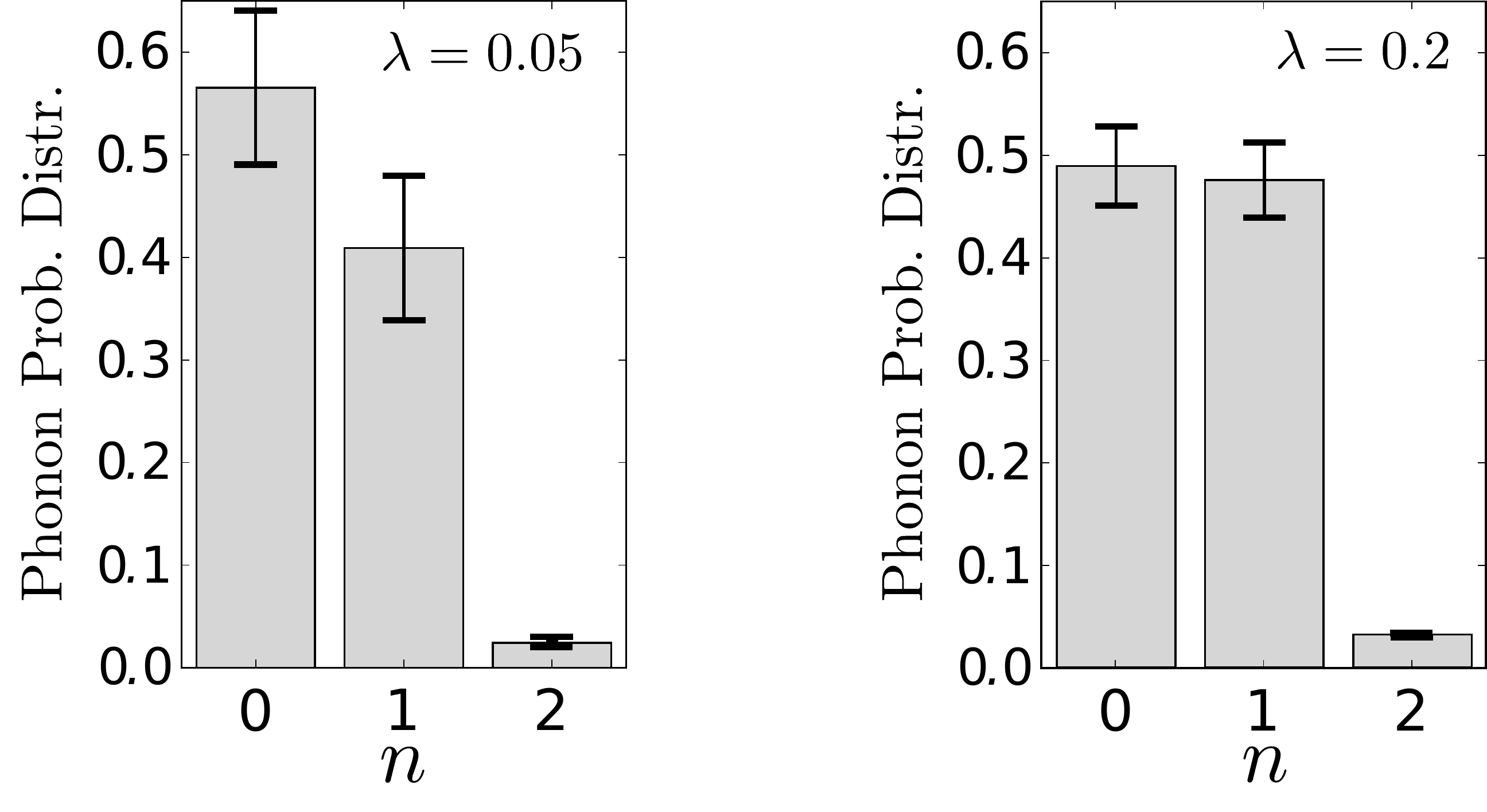}
 \caption{(Color online) Mechanical qubit preparation under non-ideal spin pre/post-selection and full decoherence dynamics. We have taken inaccuracies of $|\Delta\theta| \leq \theta \times 10^{-3}$ and $|\Delta\phi| \leq \phi \times 10^{-3}$. Other values are $t = \pi, \overline{n}_m = \overline{n}_q =10, \gamma = 10^{-3}, \Gamma = 10^{-4}, \gamma_\phi = 10^{-3}$.} \label{fig:Random}
\end{figure}

Lastly, it is important to notice that decoherence such as shown in Eq. (\ref{master2}) does not constitute the whole detrimental effects that our state preparation protocol could suffer. Although the open quantum dynamics does in fact constitutes a faithful approach towards a realistic evolution, it is important to notice that our setup rely heavily on accurate pre- and post-selection of the spin. Hence we would like to model the required accuracy in the preparation of these kind of spin states. In order to do this, we numerically compute a slight deviation of the values that fulfill the mechanical qubit preparation, i.e., a set of angles $\{\theta, \psi\}$ such as satisfy the mechanical qubit preparation condition under ideal conditions $|\lambda\eta|^2 (1-\sin\theta\cos\phi) = 1 + \sin\theta\cos\phi$. In other words, we pre- and post-select the spin state within a modest perturbation from those angles as following
\begin{equation}
 \ket{\psi_f} = \cos \left( \frac{\theta + \Delta\theta}{2}\right)\up + \sin \left( \frac{\theta + \Delta\theta}{2}\right) e^{i (\phi + \Delta\phi)}\down \label{perturbation}
\end{equation}

such as $|\Delta\theta| \ll \theta$ and $|\Delta\phi| \ll \phi$. For our understanding, this final step gives us a closer experimental realization up to where our protocol can be finally accommodated. The procedure then reads as following. Firstly, we pre-select the spin state using Eq. (\ref{perturbation}) with a random-generated distribution of $\Delta\theta$ and $\Delta\phi$. It can be seen that the ideal initial spin state considered by us throughout our manuscript was the superposition $1/\sqrt{2}(\up + \down)$, therefore the pre-selection will be considering $|\Delta\theta| \ll \pi / 2$ and $|\Delta\phi| \approx 0$. Secondly, we let the system to evolve for a time $t = \pi$ under the full master equation given in Eq. (\ref{master2}). Finally, we proceed to post-select the spin state once again using (\ref{perturbation}). In Fig. \ref{fig:Random}, we show the probability phonon distribution under both full spin-mechanical decoherence and non-ideal spin pre/post-selection. To remain within a valid mechanical preparation, we have taken an inaccuracy of 0.1$\%$ from the central values $\{\theta, \phi\}$, i.e., $|\Delta\theta| \leq \theta \times 10^{-3}$ and $|\Delta\phi| \leq \phi \times 10^{-3}$. The large standard deviation shown in the left panel of Fig. \ref{fig:Random} ($\lambda = 0.05$) can be easily understood as for this case $\langle \hat{x} \rangle$ becomes sharper/narrower at exactly the value where the coherence is maximal when $\lambda \ll 1$. Hence, an extremely accurate set of angles $\{\theta, \phi\}$ are demanded in order to prepare the mechanical qubit, a slight deviation from these central values will become in a loss of quantum coherence. To overcome this issue, the fact of having a moderate spin-mechanical coupling strength $\lambda \leq 0.25$ takes a quite relevant role when the spin post-selection accuracy is taking into account. The benefits of having stronger coupling values then arise; a deviation from the central values do not present a crucial risk to destroy the mechanical qubit preparation (as the mechanical quadratures on average become wider), and therefore the accuracy of the spin pre/post selection can be relaxed (see the right panel of Fig. \ref{fig:Random}). 

\section{Mechanical single Fock and Schr\"{o}dinger's cat preparation}\label{sec:schrodinger}
In this section, we will explore how to obtain $\ket{0}$ or $\ket{1}$ phononic states, as well as Schr\"{o}dinger's cat states through spin post-selection. The generation of Fock states for linear systems such as quantum harmonic oscillators is a challenging task to realize experimentally (see for example Refs. \cite{Hofheinz2008, Hofheinz2009, Riedinger2016-Oscillator, OConnell2010-Ground, Pepper2012-Superposition, Galland2014-Preparation, Hong2017, Chu2017}). For instance, the preparation of arbitrary photon Fock states has been experimentally achieved in superconducting quantum circuits \cite{Hofheinz2008, Hofheinz2009, Chu2017}, where a superconducting phase qubit is driven by classical microwave pulses, leading to the generation of the Fock states in a waveguide resonator.

Here, it is worthy of note that both $\ket{0}$ and $\ket{1}$ are particular solutions from (\ref{eq:condition_super}). Needless to say that, even when an initial ground state for the mechanics is considered, one would not expect this type of non-trivial solutions due to its dynamics alone. This is because each spin component displaces the mechanical state coherently by $\pm\lambda(1 - e^{-it})$, thus the election of strong (weak) enough $\lambda$ will exhibit higher (near to $\ket{0}$) phononic excitations.

\begin{figure}
 \centering \includegraphics[scale = 0.83]{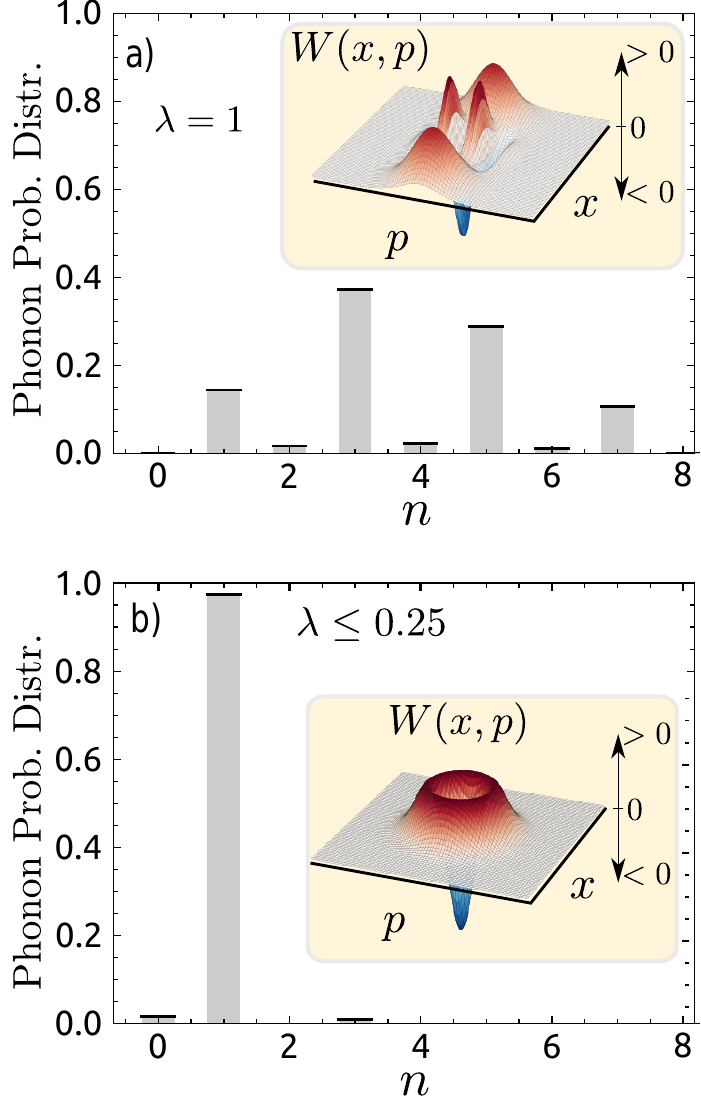}
 \caption{(Color online) a) Mechanical  Schr\"{o}dinger's cat state for $t=\pi, \phi = 0, \theta = 3\pi/2, \gamma = 0.01$ and $\lambda = 1$. b) As a consequence of decreasing the coupling, a single Fock number state $n=1$ for the same set of parameters is generated.} \label{fig:fock}
\end{figure}

From the phonon distribution shown in (\ref{prob}) in absence of any source of decoherence ($c_1 = 1,c_2 = 0$), $\mathrm{Pr}(n) \approx e^{-4\lambda^2}\lambda^{2n}(1 + \sin\theta\cos\phi (-1)^n)$, it is straightforward to see that a simple choice of the post-selected angle, e.g. $\theta = \pm 3\pi/2$ ($\phi=0$) will result into a generation of odd (or even) phonon number distribution (mechanical Schr\"{o}dinger's cat state as shown in Fig. \ref{fig:fock}-a).

In Fig. \ref{fig:fock}-b, we generate a phononic Schr\"{o}dinger's cat state working in the strong coupling regime ($\lambda \approx 1$). Interestengly, as $\lambda$ decreases ($\lambda\leq 0.25$) we can prepare the oscillator into a single Fock number state $n=1$ $(n=0)$, this being a consequence of the odd (even) phononic distribution for modest coherent amplitudes.

Finally, although we are able to prepare the mechanics into a single phonon Fock state, to post-select the qubit in the weak-coupling regime becomes hard to achieve. This could be easily seen as one proceeds to post-select the qubit into the state, e.g., $\{\theta = 3\pi/2, \phi= 0\}, \ket{\psi}_f = 1/2 (-\up + \down)$. Thus the target qubit becomes more and more orthogonal with the initial preparation. Of course, for $\lambda \rightarrow 0$ then the outcome probability $\mathcal{P}$ decreases rapidly to zero, as the mechanics disentangles from the spin. Despite of this, considering a coupling of, let us say $\lambda = 0.1$, then the Fock state $n=1$ can be prepared with a probability of $\mathcal{P}_{\mathrm{av}} = 3.8\%$ on average.

\section{Brief discussion on the AAV theory and our preparation protocol}\label{sec:aav}
At this point, we would like to stress the high resemblance of our protocol with the weak-measurement theory by Aharanov, Albert \& Vaidman (AAV) \cite{Aharonov1988-WM, Duck1989-Explanation}. Essentially, the combination of a weak interaction and pre/post-selection are shown to lead to an anomalous effect, namely the weak value amplification (WVA) \cite{Aharonov1988-WM}.

In our model one has the main ingredients of the AAV theory, hence it is interesting to study if any amplification phenomena occur. To identify this, we propose to compare the quadrature mean values obtained by post-selection measurements ($\langle \hat{x} \rangle_{\mathrm{post}}, \langle \hat{p} \rangle_{\mathrm{post}}$) with those obtained by considering the $\sigma_z$ eigenstates, i.e. $\langle \pm\lambda\eta | \hat{x} | \pm\lambda\eta \rangle = \pm\lambda(1 - \cos t), \langle \pm\lambda\eta | \hat{p} | \pm\lambda\eta \rangle = \pm \lambda \sin t$ ---similarly as in \cite{Li2014-WM}. Therefore, we define the position and momentum amplification factors as $\mathbb{Q} \equiv \langle \hat{x} \rangle_{\mathrm{post}} / 2\lambda$ and $\mathbb{P} \equiv \langle \hat{p} \rangle_{\mathrm{post}} / \lambda$, evidencing in this way, the mechanical amplification on average in our approach when $\{\mathbb{Q},\mathbb{P}\} > 1$. It can be seen from Fig. \ref{fig:coherence} that there are some cases (exactly when the mechanical qubit superposition is reached) where i) the position quadrature is amplified up to a factor of $\mathbb{Q} \approx 1/(2\times 0.05) = 10$ (and the momentum is not), ii) the momentum quadrature is amplified $\mathbb{P} \approx 1/(0.05) = 20$ (and the position is not), and iii) a combination of both. However, we cannot always identify the above amplification with the original WVA framework, since in our case the AAV approximation breaks down \cite{Duck1989-Explanation} (see Appendix \ref{app:Aharanov} for more details).

\section{Concluding remarks}\label{sec:conclusion}
We have proposed a feasible probabilistic method to generate a macroscopic mechanical qubit, as well as Schr\"{o}dinger's cat and single Fock number states ($n=1$) for the oscillator. As opposed to previous works \cite{Heeres2015-Nonlinear, Krastanov2015-Pumping, Khan2016-Optomechanics}, we studied an open dispersive spin-mechanical system without any spin and/or mechanical driving, but on non-ideal spin post-selection measurements in the weak/moderate coupling regime.

To understand the mechanical qubit superposition, we derive a correspondence between the amplification of the mechanical quadratures on average and the maximum value of the mechanical quantum coherence ---whereas the single Fock number state is a direct consequence of low-amplitude Schr\"{o}dinger's cat states.

From an experimental point of view, our scheme shows to be feasible under current technology, as we demonstrated in Sec. \ref{sec:master}-B where our scheme can be accomodated in presence of several sources of decoherence. Moreover, our technique has a probability of production of $2\% < \mathcal{P}_{\mathrm{av}} <  24\%$ in the range of $0.05 < \lambda < 0.2$, respectively. 

Lastly, as our protocol is mainly based on spin post-selection, we would like to illustrate how this procedure could be addressed. Firstly, in order to pre-select the spin qubit, we initialize the spin in the $\up$ state, subsequently we apply a $(\pi/2)_y$ rotation around y-axis that prepares the spin into a coherent superposition. Secondly, on the spin post-selection, we apply a $(\theta/2)_y$ rotation ($\theta$ is the post-selected angle) to map the coherent superposition into $\hat{\sigma}_z$ eigenstates, and then a measure on the spin $\up$ component will allow us to read-out the desired post-selected state.

\section*{ACKNOWLEDGMENTS}
V.M. and R.C acknowledge the financial support of the projects Fondecyt Postdoctorado $\#$3160700 and $\#$3160154, respectively. M.O. and V.E. acknowledge the financial support of the project Fondecyt $\#$1140994. We have been greatly benefited from comments of S. Bose, J. Maze, L. Davidovich, R. de Matos Filho, N. Zagury and F. de Melo.

\appendix

\section{Failure of the mechanical quadrature mean value under weak-measurements}\label{app:Aharanov}

Aharanov, Albert and Vaidman's (AAV) theory rapidly attained serious interests and debate since its conception in the late 1980s \cite{Aharonov1988-WM}. In the original paper the attention is centered in a new paradigm of quantum measurements, where a combination of a weak interaction followed by a strong (projective) measurement could lead to an anomalous effect, namely the mean values of physical observables go beyond the eigenvalue spectrum. To illustrate this, the seminal work considered a spin-1/2 particle passing through two Stern-Gerlach apparatus. The first one is aimed to pre-select the spin-1/2 state via weak magnetic coupling (weak enough not to perturb significantly the quantum state), whereas the second one will post-select the particle state. Surprisingly, the measurement result of the spin component could lie way beyond its eigenvalue spectrum.

Commonly in AAV's framework, one is interested in quantifying the quadratures on average of the mechanical/meter state as they are closely linked to the weak-value amplification. In the main text, as pointed out by us, although our protocol has some of the ingredients of weak-measurements, it breaks down shortly after considering our optimal values of coupling strength ($\lambda$) and post-selection values ($\{\theta, \phi\}$).

In this section, we will explicitly show how the mechanical quadratures on average differ when these are computed using standard rules from quantum mechanics ---where no approximations on the coupling strength have been done--- and the one following the weak measurements approximations \cite{Aharonov1988-WM, Duck1989-Explanation}. In order to do this, let us recall the Hamiltonian:
\begin{equation}
 \hat{H}_{\mathrm{int}} = \hat{b}^\dagger \hat{b} - \lambda \hat{\sigma}_z (\hat{b}^\dagger + \hat{b}),
\end{equation}

with unitary evolution operator
\begin{equation}
 \hat{U} = e^{\lambda \hat{\sigma}_z \hat{\Lambda}}e^{-i\hat{b}^\dagger\hat{b} t}.
\end{equation}

In the above, we have defined $\hat{\Lambda} \equiv \eta \hat{b}^\dagger - \eta^* \hat{b}$, and $\eta \equiv 1 - e^{-i t}$. The initial state reads as $\ket{\psi(0)} = \ket{\psi_i} \otimes \ket{\alpha_m}$, where $\ket{\psi_i} = 1/\sqrt{2}(\up + \down)$ and $\ket{\alpha_m}$ is the mechanical coherent state. Following with the same procedure of the AAV's theory, we proceed to approximate the unitary operator at first order in the coupling strength as
\begin{equation}
 \ket{\psi(t)} \approx \left(\mathbb{I} + \lambda \hat{\sigma}_z \hat{\Lambda}\right) \ket{\psi_i} \otimes \ket{\alpha_m e^{-it}}.
\end{equation}

This approximation remains valid if and only if $\lambda \ll 1$. To obtain the relevant mechanical state, we post-select the spin giving us the following wave-function:
\begin{equation}
 \ket{\psi_f}\bra{\psi_f} \psi(t)\rangle = \bra{\psi_f}\psi_i\rangle ( \mathbb{I} + \lambda \langle \hat{\sigma}_z \rangle_w \hat{\Lambda})\ket{\psi_f}\otimes\ket{\alpha_m e^{-i t}},
\end{equation}

where the weak value is defined by:
\begin{equation}
 \langle \hat{\sigma}_z \rangle_w \equiv \frac{\bra{\psi_f}\hat{\sigma}_z\ket{\psi_i}}{\bra{\psi_f}\psi_i\rangle} = \mathcal{A} + i \mathcal{B}.
\end{equation}

\begin{figure}[t]
 \centering \includegraphics[scale = 0.4]{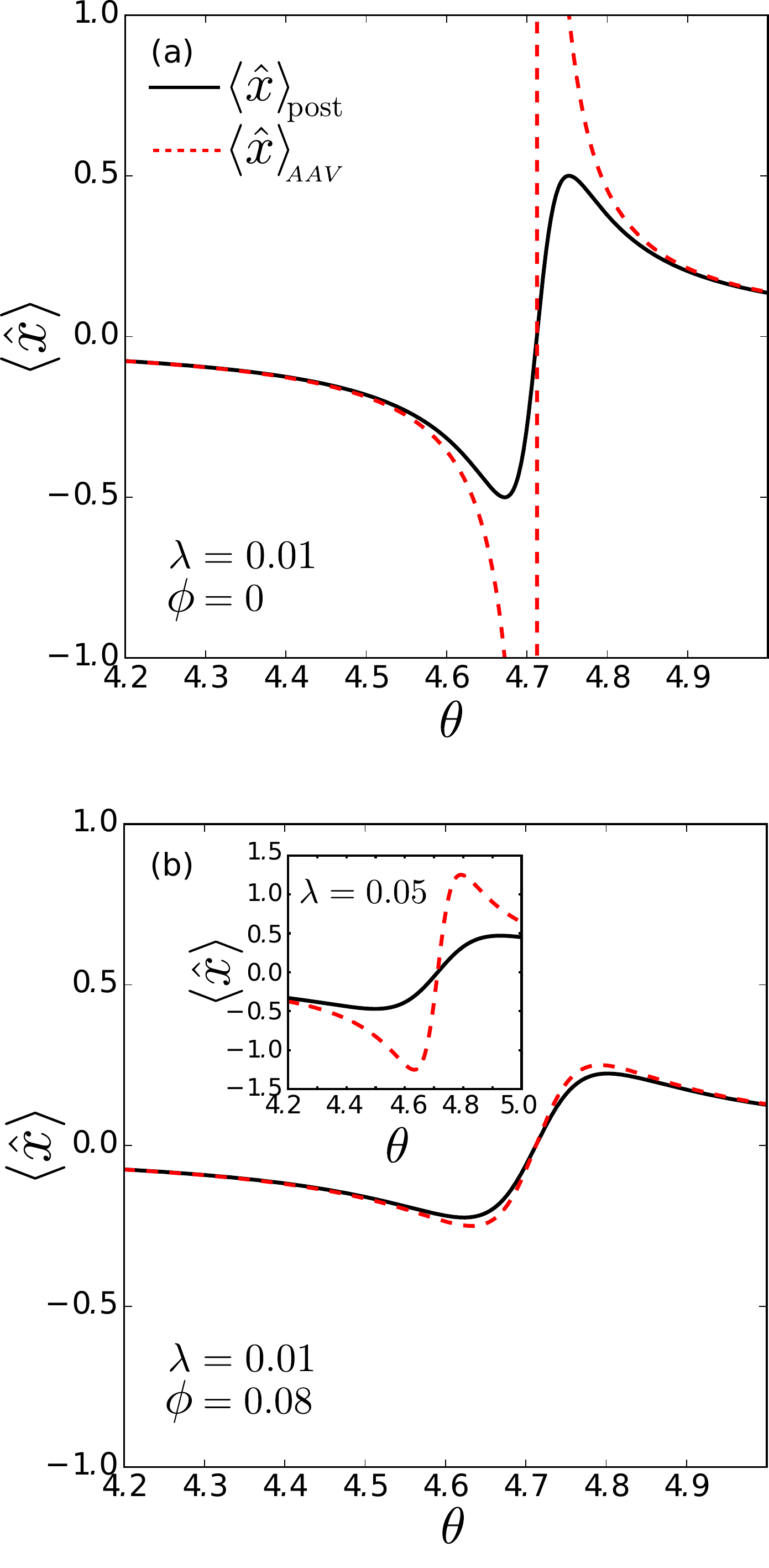}
 \caption{(Color online) Comparison between the position of the mechanical object on average calculated as in the main text (no approximations considered) (black solid line), and the one using weak-measurement approximation (red dashed line) (see Eq. (\ref{eq:wv})).}
 \label{fig:wva}
\end{figure}

From the above, the unnormalized mechanical state reads as 
\begin{equation}
\ket{\alpha} = \left(\mathbb{I} + \lambda  \langle \hat{\sigma}_z \rangle_w \hat{\Lambda}\right) \ket{\alpha_m e^{-it}}. \label{mechanicalstate}
\end{equation}

On the other hand, let us consider $\hat{M}$ a quantum observable of the mechanical object. Therefore, its expectation value is computed as usual
\begin{eqnarray}
 \langle \hat{M} \rangle &=& \frac{\bra{\alpha} \hat{M} \ket{\alpha}}{\bra{\alpha}\alpha\rangle}\\
 \nonumber  &=&  \langle \hat{M} \rangle_0 + \lambda (\mathcal{A} \langle [\hat{M},\hat{\Lambda}]\rangle_0 + i \mathcal{B} \langle \{\hat{M},\hat{\Lambda}\}\rangle_0)\\
  &-& 2 i \lambda \mathcal{B} \langle \hat{M} \rangle_0\langle\hat{\Lambda}\rangle_0).\label{m}
\end{eqnarray}

where $\langle \hat{O} \rangle_0 \equiv \langle \alpha_m e^{-i t}| \hat{O} | \alpha_m e^{-i t}\rangle$. We are interested to contrast our results from the main work ($\langle\hat{x}\rangle = \mathrm{Tr}[\hat{\rho}\hat{x}]$, where $\hat{\rho}$ is the mechanical state after the spin post-selection without any approximation) with the one presented in Eq. (\ref{m}). Thus, let us consider $\hat{M} = \hat{x} = (\hat{b} + \hat{b}^\dagger)/2$. In this case $[\hat{x}, \hat{\Lambda}] = 1 - \cos t, \langle \{\hat{x}, \hat{\Lambda}\} \rangle = i \left( (1 + 2 |\alpha_m|^2)\mathrm{Im}[\eta] + 2 \mathrm{Im}[\eta \alpha_m^{*2}e^{2 i t}] \right), \langle \hat{x} \rangle = \mathrm{Re}[\alpha_m e^{-i t}],$ and $\langle \hat{\Lambda} \rangle = 2 i \mathrm{Im}[\eta \alpha_m^* e^{i t}]$ giving the final expression:

\begin{eqnarray}
 \nonumber \langle \hat{x} \rangle_{AAV} &=& \mathrm{Re}\left[\alpha_m e^{-i t}\right] + \lambda \mathcal{A} (1 - \cos t)\\
 \nonumber &-& \lambda \mathcal{B}\left\{ (1 + 2 |\alpha_m|^2)\mathrm{Im}[\eta] + 2 \mathrm{Im}[\eta \alpha_m^{*2}e^{2 i t}] \right\}\\
 &+& 4 \lambda \mathcal{B} \mathrm{Re}\left[\alpha_m e^{-i t}\right] \mathrm{Im}\left[\eta \alpha_m^{*} e^{i t}\right]. \label{eq:wv}
\end{eqnarray}

In particular, we took the mechanical state initialized in its ground state throughout our work. Hence $\alpha_m = 0$, and the mean value is then reduced to:
\begin{equation}
 \langle \hat{x} \rangle_{AAV} = \lambda \left\{ \mathcal{A} (1 - \cos t) - \mathcal{B}\sin t  \right\}.
\end{equation}

In Fig. \ref{fig:wva} we illustrate the position mean value of the mechanical oscillator as a function of the post-selected angles $\{\theta, \phi\}$. In the top panel of Fig. \ref{fig:wva} we consider the case where $\phi = 0$. As expected, if $\phi = 0$ and $\theta \rightarrow 3\pi/2$ then $\langle \hat{\sigma}_z \rangle_w \rightarrow \infty$. Therefore, Eq. (\ref{mechanicalstate}) becomes undefined. Of course, at exactly the angle value of $\theta = 3\pi/2$ the weak-measurement approximation is not be valid, as the original paper forbid orthogonal spin post-selection related to the initial spin state. However, one should expect that in the vicinity of $\theta \rightarrow 3\pi/2$, the weak-measurement approximation should hold. As seen in Fig. \ref{fig:wva}-a and discussed in the main text, when $\mathrm{max}(\langle \hat{x} \rangle_{\mathrm{post}})$ (or $\mathrm{min}(\langle \hat{x} \rangle_{\mathrm{post}})$) occurs, the mechanical qubit state is prepared. However the weak-measurement approximation becomes irreconcilable with the exact calculation for an independent choice of the coupling value $\lambda$. On the other hand, our protocol holds for any set of $\{\theta, \phi\}$ fulfilling $|\lambda\eta|^2(1-\sin\theta\cos\phi) = 1+\sin\theta\cos\phi$, including $\phi = 0$ as shown in Fig. \ref{fig:wva}-a.

To explore the validity of the weak-measurement approximation when $\phi \neq 0$ (no-orthogonal states between spin pre- and post-selection), we have calculated $\langle \hat{x} \rangle$ for $\phi = 0.08$ (see Fig. \ref{fig:wva}-b). There, the approximation is valid for $\lambda \leq 0.01$. However, as discussed before, an experimental drawback of having $\lambda \ll 1$ is that we require to achieve extremely high precision in the spin post-selection, and also the spin outcome probability decreases. Thus, even though when $\phi \neq 0$ and $\lambda \ll 1$ the AAV is valid, the experimental disadvantages become in a major obstacle for the mechanical qubit preparation.

\bibliographystyle{apsrev4-1}
\bibliography{Mechanical_Qubit_PRA}

\begin{thebibliography}{81}
\bibitem [{GS-()}]{GS-Justification}%
  \BibitemOpen
  \href@noop {} {\ }\bibinfo {note} {As we are interested in the generation of
  a mechanical qubit state, we require to have low initial mechanical quanta
  excitations for our protocol to happen. Hence, even though we have considered
  a coherent mechanical amplitude of $\alpha = 0$ (ideal case) throughout our
  work, numerical simulations show that it could be slightly relaxed up to
  values of $|\alpha| \leq 0.3$. Of course, when $|\alpha|$ increases it will
  be at expenses of having higher mechanical phononic states, thus destroying a
  perfect mechanical qubit preparation.}\BibitemShut {Stop}%
\bibitem [{exp()}]{explanationplussign}%
  \BibitemOpen
  \href@noop {} {\ }\bibinfo {note} {The target mechanical qubit state with
  minus sign $\ket{\psi_m} = 1/\sqrt{2} (\ket{0} - \ket{1})$ can also be
  obtained with our state preparation protocol. In fact, it can be seen
  straighforwardly in the main text (see Fig. (2) therein) corresponding to the
  minimum value of the position quadrature on average. For our purpose, we only
  considered the case where we are able to prepare $\ket{\psi_m} = 1/\sqrt{2}
  (\ket{0} + \ket{1})$, i.e., the maximal value of $\langle \hat{x}
  \rangle$.}\BibitemShut {Stop}%
\end{thebibliography}

\end{document}